\documentstyle[12pt,citesort,subeqn]{ioplppt}
\begin{document}
\makeatletter
\@addtoreset{equation}{section}
\def\theequation{\thesection.\arabic{equation}}
\makeatother
\def\Mae{\mbox{\hspace*{-20pt}}}
\def\Cmae{\mbox{\hspace*{-10pt}}}
\def\Usr{\mbox{\hspace*{20pt}}}
\def\Cusr{\mbox{\hspace*{10pt}}}
\def\Define{\mathop{\stackrel{\rm def}{=}}}
\def\ledo{{\stackrel{\rm D}{\leq}}}
\newcommand{\myref}[1]{(\ref{#1})}
\newcommand{\vecvar}[1]{\mbox{\boldmath$#1$}}
\title{Symmetric Fock space and orthogonal symmetric polynomials
associated with the Calogero model}[Symmetric Fock space and the
Calogero model]
\author{Akinori Nishino%
\footnote{E-mail address: nishino@monet.phys.s.u-tokyo.ac.jp},
Hideaki Ujino\footnote{E-mail address: ujino@monet.phys.s.u-tokyo.ac.jp}
and Miki Wadati\footnote{E-mail address: %
wadati@monet.phys.s.u-tokyo.ac.jp}\\
Department of Physics, Graduate School of Science,\\
University of Tokyo,\\
Hongo 7--3--1, Bunkyo-ku, Tokyo 113--0033, Japan}
\date{}
\maketitle


\begin{abstract}
Using a similarity transformation that maps the Calogero model into $N$
decoupled quantum harmonic oscillators, we construct a set of mutually
commuting conserved operators of the model and their simultaneous
eigenfunctions.
The simultaneous eigenfunction is a deformation of the symmetrized number
state (bosonic state) and forms an orthogonal basis of the Hilbert (Fock)
space of the model. This orthogonal basis is different from the
known one that is a variant of the Jack polynomial, i.e.,
the Hi-Jack polynomial.
This fact shows that the conserved operators derived by the similarity
transformation and those derived by the Dunkl operator formulation
do not commute. Thus we conclude that the Calogero model has two,
algebraically inequivalent sets of mutually commuting conserved
operators, as is the case with the hydrogen atom. We also confirm
the same story for the $B_{N}$-Calogero model.
\end{abstract}


\section{Introduction}
Exact solutions for the Schr\"{o}dinger equations have provided interesting
problems in physics and mathematical physics since the early days of quantum
mechanics. Special functions such as the Hermite polynomials and the Laguerre
polynomials play important roles in the study of the quantum harmonic
oscillator and the hydrogen atom. Such a traditional approach to the
quantum mechanics is enjoying a revived interest since the importance of
the Jack symmetric polynomials was realized in the calculation of correlation
functions of the Sutherland model~\cite{Sutherland1,Sutherland2,Jack,%
Macdonald,Stanley,Forrester1,Forrester2,Ha1,Ha2}.
The Sutherland model is one of the typical models
among the one-dimensional quantum integrable systems with inverse-square
long-range interactions that has been extensively studied with the
helps of the Dunkl operator formulation and the theory of the Jack polynomials.
Even an exact calculation of its dynamical density-density correlation
function~\cite{Ha1,Ha2} and
an algebraic construction of its orthogonal basis~\cite{Lapointe1,Lapointe2},
i.e., the Jack polynomials, have been achieved. 

The Calogero model is another typical model among the class~\cite{Calogero}.
Its Hamiltonian is given by 
\begin{equation} 
  \hat{H}_{\rm C}^{(A)}
  =\frac{1}{2}\sum_{j=1}^{N}(p_{j}^{2}+\omega^{2}x_{j}^{2}) 
           +\frac{1}{2}\sum_{\stackrel{\scriptstyle j,k=1}{j\neq k}}^{N}
           \frac{a(a-1)}{(x_{j}-x_{k})^{2}}, 
  \label{A-calogero1} 
\end{equation}
where the constants $a$ and $\omega$ are the coupling parameter and the 
strength of the external harmonic well, respectively,
and $p_{j}$ is a momentum operator,
$p_{j}=-\i\frac{\partial}{\partial x_{j}}$.
We note that the coordinate representation, or identification of the
momentum with the partial differential operator, is implicitly assumed
throughout the paper. The mass of the particles and the Planck constant
$\hbar$ are taken to be unity.  Strictly speaking the model 
\myref{A-calogero1} is introduced by Sutherland~\cite{Sutherland3}. 
Calogero originally introduced the model
with harmonic interactions, which is obtained from the model with the
harmonic well by fixing the center of mass at the coordinate origin
as is presented in \myref{A-calogero1}.
The superscript $(A)$ on the Hamiltonian means that it is invariant under
the action of the  $A_{N-1}$-type Weyl group, i.e.\ under $S_{N}$,
on the indices of the particle.
Thus the model is sometimes called the $A_{N-1}$-Calogero model%
~\cite{Olshanetsky1}.
Because of its structural similarity to the quantum harmonic oscillator,
several ways of algebraic construction of the eigenfunctions of the
Calogero Hamiltonian had been demonstrated~\cite{Perelomov1,Brink1,Brink2,%
Ujino0,Ujino1} before the Rodrigues formula for the Jack polynomials
appeared. However, identification of its orthogonal basis was missing
for a long time~\cite{Ujino2}. Motivated by the Rodrigues formula for the Jack
polynomials, we derived the Rodrigues formula for the Hi-Jack
(or multivariable Hermite) polynomials for the first
time~\cite{Ujino3,Ujino4}, which is now identified with an
orthogonal basis for the Calogero model~\cite{Ujino5,Baker,vanDiejen}.
These studies mentioned above have stimulated
lots of works on variants of Jack polynomials and integrable systems
with inverse-square interactions.
The multivariable Laguerre polynomials associated with the 
$B_{N}$-type Calogero
model,
\begin{equation}
  \label{B-calogero1}
  \fl \hat{H}_{\rm C}^{(B)} = \frac{1}{2}\sum_{j=1}^{N}
  \left(p_{j}^{2}+\omega^{2}x_{j}^{2}
  +\frac{b(b-1)}{x_{j}^{2}}\right)
  +\frac{1}{2}\sum_{\stackrel{\scriptstyle j,k=1}{j\neq k}}^{N}
  \left(\frac{a(a-1)}{(x_{j}-x_{k})^{2}}
  +\frac{a(a-1)}{(x_{j}+x_{k})^{2}}\right),
\end{equation}
where a constant $b$ is another coupling parameter besides that of the
$A_{N-1}$ case, has attracted lots of interest as such a variant~\cite{%
Baker,vanDiejen,Kakei1}.
The superscript $(B)$ on the Hamiltonian means that it is invariant under
the action of the  $B_{N}$-type Weyl group on the indices of the particle.
Through the Dunkl operator formulations
for the above three models~\cite{Dunkl,Polychronakos,Yamamoto},
we realize that
the Jack, Hi-Jack and multivariable Laguerre polynomials are the
simultaneous eigenfunctions of the conserved operators of the
corresponding models~\cite{Ujino5,Kakei1}.
See \ref{sec:Dunkl-f} for detail.

Quite recently, Gurappa and Panigrahi presented similarity transformations
that map the Calogero and $B_{N}$-Calogero models into $N$ decoupled
quantum harmonic oscillators~\cite{Gurappa1,Gurappa2}.
Their transformation for the $A_{N-1}$ case is,
in a sense, equivalent to a transformation to the Euler operator,
$\sum_{j=1}^{N}x_{j}\frac{\partial}{\partial x_{j}}$, which
had been shown by Sogo~\cite{Sogo}.
Reformulating their results,
we noticed the connection of the number operator
and the symmetrized number state (bosonic state) of the harmonic
oscillators with sets of conserved operators and
symmetric orthogonal bases of the $A_{N-1}$- and $B_{N}$-Calogero models%
~\cite{Ujino6}. The purpose of this paper is to present detailed
properties of the orthogonal basis derived by the similarity transformation
method.

The outline of the paper is as follows.
In section \ref{sec:similarity_transform},
we present similarity transformations from the Calogero models to
the decoupled harmonic oscillators. In section \ref{sec:new-orthogonal},
we construct
a set of conserved operators from the number operators and their
simultaneous eigenfunctions
that form orthogonal bases of the models.
Some properties of the simultaneous eigenfunctions are presented.
In section \ref{sec:relation},
we compare the new orthogonal bases with the known
orthogonal bases. And we show that the conserved operators constructed by
the similarity transformation method and those constructed by the Dunkl
operator formulation are algebraically inequivalent.
In section \ref{sec:summary},
we summarize the results and discuss future problems.
\ref{sec:Dunkl-f} presents a brief summary on the Dunkl operator formulation
and variants of the Jack polynomial. \ref{sec:singularity} covers
detailed discussions on the cancellation of essential singularities.
\ref{sec:explicit_forms} shows explicit forms
of some of the new orthogonal basis.


\section{Similarity transformation to harmonic oscillators}
\label{sec:similarity_transform}
We show a series of similarity transformations that map the 
Hamiltonians of the 
Calogero models to that of the $N$ decoupled quantum 
harmonic oscillators~\cite{Gurappa1,Gurappa2}.
The ground state wave function $\Psi_{\rm g}^{(A)}$ 
and the ground state energy $E_{\rm g}^{(A)}$
for the $A_{N-1}$-Calogero model are given by 
\begin{equation}
  \label{A-gswf}
  \Psi_{\rm g}^{(A)}(\mbox{\boldmath $x$})
  =\prod_{1\leq j<k\leq N}|x_{j}-x_{k}|^{a}\,
  \exp\left( -\frac{1}{2}\omega\sum_{l=1}^{N}x_{l}^2 \right),
\end{equation}
\begin{equation}
  \label{A-gserg}
  E_{\rm g}^{(A)}=\frac{1}{2}\omega N\bigl((N-1)a+1\bigr).
\end{equation}
An excited state is written by a product of the ground state and
some symmetric polynomial $\phi(\mbox{\boldmath $x$})$,
$\Psi^{(A)}=\phi(\mbox{\boldmath $x$})\Psi_{\rm g}^{(A)}$.
Since we are interested in the symmetric polynomial part
of the excited state, we perform a similarity 
transformation and remove the ground state from the operand of
the Hamiltonian (\ref{A-calogero1}), 
\begin{eqnarray}
  \label{A-calogero2}
  \fl H_{\rm C}^{(A)} \Define (\Psi_{\rm g}^{(A)})^{-1}
  (\hat{H}_{\rm C}^{(A)}-E_{\rm g}^{(A)})\Psi_{\rm g}^{(A)}
  \nonumber \\ 
  \fl \Usr \; = \sum_{l=1}^{N}\left(-\frac{1}{2}
  \frac{\partial^{2}}{\partial x_{l}^{2}}
  +\omega x_{l}\frac{\partial}{\partial x_{l}}\right)
  -\frac{1}{2}a\sum_{\stackrel{\scriptstyle l,m=1}{l\neq m}}^{N}
  \frac{1}{x_{l}-x_{m}}\left(\frac{\partial}{\partial x_{l}}
  -\frac{\partial}{\partial x_{m}}\right).
\end{eqnarray}
We apply a similar procedure to the $B_{N}$-Calogero model
\myref{B-calogero1}.
Using the ground state wave function $\Psi_{\rm g}^{(B)}$
and the ground state energy $E_{\rm g}^{(B)}$
for the $B_{N}$-Calogero 
model,
\begin{equation}
  \label{B-gswf}
  \fl \Psi_{\rm g}^{(B)}(\mbox{\boldmath $x$})
  =\prod_{1\leq j<k\leq N}|x_{j}-x_{k}|^{a}|x_{j}+x_{k}|^{a}
   \prod_{l=1}^{N}|x_{l}|^{b}
   \exp\left(-\frac{1}{2}\omega\sum_{m=1}^{N}x_{m}^{2}\right),
\end{equation}
\begin{equation}
  \label{B-gserg}
  \fl E_{\rm g}^{(B)}=\frac{1}{2}\omega N\bigl(2(N-1)a+2b+1\bigr),
\end{equation}
we transform the $B_{N}$-Calogero model as
\begin{eqnarray}
  \label{B-calogero2}
  \fl H_{\rm C}^{(B)} \Define (\Psi_{\rm g}^{(B)})^{-1}
  (\hat{H}_{\rm C}^{(B)}-E_{\rm g}^{(B)})\Psi_{\rm g}^{(B)}
  \nonumber \\
  \fl \Cusr = \sum_{l=1}^{N}\left(
  -\frac{1}{2}\frac{\partial^{2}}{\partial x_{l}^{2}}
  -\frac{b}{x_{l}}\frac{\partial}{\partial x_{l}}
  +\omega x_{l}\frac{\partial}{\partial x_{l}}\right)
  -a\sum_{\stackrel{\scriptstyle l,m=1}{l\neq m}}^{N}
  \frac{1}{x_{l}^{2}-x_{m}^{2}}
  \left(x_{l}\frac{\partial}{\partial x_{l}}
  -x_{m}\frac{\partial}{\partial x_{m}}\right).
\end{eqnarray}
Though the discussions above seem to be restricted to the bosonic wave
functions, they cover any choice of statistics.
The statistics of the particles, or in other words, the symmetry of the
wave functions of the Calogero model is determined by a choice of the
phase of the Jastraw factor $\prod_{1\le j<k\le N}|x_{j}-x_{k}|^{a}$
in the ground state wave function. In fact, we can choose any phase.
For instance, a function,
\begin{equation}
 \prod_{1\le j<k\le N}|x_{j}-x_{k}|^{a}({\rm sgn}(x_{j}-x_{k}))^{m}
 \exp\left(-\frac{1}{2}\omega\sum_{l=1}^{N}x_{l}^{2}\right),
 \label{}
\end{equation}
where $0\le m<2$, is the ground state of the $A_{N-1}$-Calogero model.
This fact is related to impenetrability of the inverse square potential
in one-dimension~\cite{Sutherland3,Calogero}. There must be some deep
physical meaning behind the fact, which is beyond the scope of this paper.
We note that the $B_{N}$-Calogero model also has similar freedom in the
choice of the phase of the ground state wave function. The choice of the
phase has no effect in the following study. Thus we have taken the
simplest choice as a representative.
In what follows, we sometimes call the operators \myref{A-calogero2} and
(\ref{B-calogero2}) Hamiltonians of the $A_{N-1}$- and $B_{N}$-Calogero
models instead of the original Hamiltonians \myref{A-calogero1}
and \myref{B-calogero1}.

Introducing the Euler operator ${\cal O}_{{\rm E}}$ and the Lassalle
operators ${\cal O}_{{\rm L}}^{(A,B)}$~\cite{Lassalle1,Lassalle2},
we can rewrite the Hamiltonians of the $A_{N-1}$- 
and $B_{N}$-Calogero models in a unified fashion,
\begin{equation}
  \label{AB-calogero3}
  H_{{\rm C}}^{(A,B)}=\omega{\cal O}_{{\rm E}}
                          -\frac{1}{2}{\cal O}_{{\rm L}}^{(A,B)},
\end{equation}
where
\begin{subequations}
  \begin{eqnarray}
  \fl {\cal O}_{{\rm E}}
      \Define\sum_{l=1}^{N}x_{l}\frac{\partial}{\partial x_{l}}, \\ 
  \fl {\cal O}_{{\rm L}}^{(A)}\Define
      \sum_{l=1}^{N}\frac{\partial^{2}}{\partial x_{l}^{2}} 
      +a\sum_{\stackrel{\scriptstyle l,m=1}{l\neq m}}^{N}
      \frac{1}{x_{l}-x_{m}}
      \left(\frac{\partial}{\partial x_{l}}
          -\frac{\partial}{\partial x_{m}}\right), \\ 
  \fl {\cal O}_{{\rm L}}^{(B)}\Define
      \sum_{l=1}^{N}\left(\frac{\partial^{2}}{\partial x_{l}^{2}}
      +\frac{2b}{x_{l}}\frac{\partial}{\partial x_{l}}\right)
      +2a\sum_{\stackrel{\scriptstyle l,m=1}{l\neq m}}^{N}
      \frac{1}{x_{l}^{2}-x_{m}^{2}}
      \left(x_{l}\frac{\partial}{\partial x_{l}}
      -x_{m}\frac{\partial}{\partial x_{m}}\right).
\end{eqnarray}
\end{subequations}
Since the commutation relations between the Euler operator 
${\cal O}_{{\rm E}}$ 
and the Lassalle operators ${\cal O}_{{\rm L}}^{(A,B)}$ are 
\begin{equation}
  \label{E-L-commutation}
  \bigl[{\cal O}_{{\rm L}}^{(A,B)},{\cal O}_{{\rm E}}\bigr]
  =2{\cal O}_{{\rm L}}^{(A,B)},
\end{equation}
both Hamiltonians have the same algebraic structure.
For a while, we omit the superscript $(A)$ and $(B)$ to avoid
complexity and duplication of the expressions.
Through \myref{E-L-commutation} and the Baker-Hausdorff
formula, we confirm that
the Hamiltonians are transformed into the Euler operator~\cite{Sogo}, 
\begin{equation}
  \label{transform1}
  \e^{\frac{1}{4\omega}{\cal O}_{{\rm L}}}H_{\rm C}
  \e^{-\frac{1}{4\omega}{\cal O}_{{\rm L}}}=\omega {\cal O}_{{\rm E}},
\end{equation}
which gives decompositions of the two models into the total momentum
operator for $N$ interaction-free particles on a ring of 
circumference $L$ with the identification 
$x_{j}=\exp\frac{2\pi\i}{L}\theta_{j}$.

Furthermore, we transform the Euler operator into the Hamiltonian 
of the decoupled quantum harmonic oscillators.
The following commutation relations,
\begin{equation}
  \label{Laplacian}
  \bigl[\triangle,{\cal O}_{{\rm E}}\bigr]=2\triangle, \quad 
  \bigl[\mbox{\boldmath $x$}^{2},{\cal O}_{{\rm E}}\bigr]
   =-2\mbox{\boldmath $x$}^{2}, \quad 
  \bigl[\triangle,\mbox{\boldmath $x$}^{2}\bigr]=2(2{\cal O}_{{\rm E}}+N), 
\end{equation}
where the symbols $\triangle$ and $\mbox{\boldmath $x$}^{2}$ denote
the Laplacian and the square of the norm,
\begin{equation}
  \triangle\Define
  \sum_{j=1}^{N}\frac{\partial^{2}}{\partial x_{j}^{2}}, \quad \quad 
  \mbox{\boldmath $x$}^{2}\Define\sum_{j=1}^{N}x_{j}^{2},
\end{equation}
and again the Baker-Hausdorff formula yield 
\begin{equation}
  \label{transform2}
  \e^{-\frac{1}{4\omega}\triangle} 
   \omega{\cal O}_{{\rm E}} 
  \e^{\frac{1}{4\omega}\triangle} 
  =\omega{\cal O}_{{\rm E}}-\frac{1}{2}\triangle.
\end{equation}
Finally, the similarity transformation using the Gaussian kernel produces
the Hamiltonian of 
the $N$ decoupled quantum 
harmonic oscillators with its ground state energy subtracted from it,
\begin{equation}
  \label{transform3}
  \e^{-\frac{1}{2}\omega\mbox{\boldmath $x$}^{2}} 
  \e^{-\frac{1}{4\omega}\triangle}
   \omega{\cal O}_{{\rm E}} 
  \e^{\frac{1}{4\omega}\triangle} 
  \e^{\frac{1}{2}\omega\mbox{\boldmath $x$}^{2}} 
  = \frac{1}{2}\sum_{j=1}^{N}
  (p_{j}^{2}+\omega^{2}x_{j}^{2})-\frac{1}{2}N\omega .
  \label{eqn:harmonic_oscillator}
\end{equation}
In terms of the creation and annihilation operators of the quantum 
harmonic oscillators, 
\begin{subequations}
\begin{eqnarray}
  \label{creation-annihilation}
  a_{j}^{\dagger}=\frac{1}{2\omega\i}(p_{j}+\i\omega x_{j}), \\
  a_{j}=\i(p_{j}-\i\omega x_{j}), \\
  n_{j}=a_{j}^{\dagger}a_{j}
  =\frac{1}{2\omega}(p_{j}^{2}+\omega^{2}x_{j}^{2})-\frac{1}{2},
\end{eqnarray}
\end{subequations}
the r.h.s. of (\ref{eqn:harmonic_oscillator}) becomes the sum
of the number operators, $\omega\sum_{j=1}^{N}n_{j}$.
To summarize, we get the similarity transformations,
\begin{equation}
  \label{transform4}
  T^{-1}H_{\rm C}T=\omega\sum_{j=1}^{N}n_{j},\quad
  T\Define\e^{-\frac{1}{4\omega}{\cal O}_{{\rm L}}}
               \e^{\frac{1}{4\omega}\triangle} 
               \e^{\frac{1}{2}\omega\mbox{\boldmath $x$}^{2}},
\end{equation}
which map the $A_{N-1}$- and $B_{N}$-Calogero Hamiltonians
to that of the 
$N$ decoupled quantum harmonic oscillators. The number operators, 
$n_j$, $j=1,2,\cdots,N$, are mutually
commuting conserved operators of the quantum harmonic
oscillators. Their
non-degenerate simultaneous eigenfunctions are
nothing but the (nonsymmetric) number states, 
\begin{equation}
  \label{number-state}
  |n_{1},\cdots,n_{N}\rangle
  \Define\prod_{j=1}^{N}(a_{j}^{\dagger})^{n_{j}}|0\rangle,
\end{equation}
where $|0\rangle\Define\e^{-\frac{1}{2}\omega\vecvar{x}^{2}}$
is the vacuum state for the quantum harmonic oscillators. Note that we
implicitly employ the coordinate representation, 
$|0\rangle\sim\langle\vecvar{x}|0\rangle$.
We are tempted to conclude that the similarity transformation of
the number state $T|n_{1},\cdots,n_{N}\rangle$
back to the Hilbert space of the Calogero models
gives the non-symmetric orthogonal basis of the models.
However, as we shall see in the next section, this conclusion is wrong
because of a specific property of the Lassalle operators.


\section{Conserved operators and orthogonal bases}
\label{sec:new-orthogonal}
Let us consider the similarity transformation from the number state back to
the Hilbert space of the Calogero models. It is easy to verify,
\begin{subequations}
  \begin{eqnarray}
    & & x_{j} =
    {\rm e}^{\frac{1}{4\omega}\triangle}
    {\rm e}^{\frac{1}{2}\omega\vecvar{x}^{2}}a_{j}^{\dagger}
    {\rm e}^{-\frac{1}{2}\omega\vecvar{x}^{2}}
    {\rm e}^{-\frac{1}{4\omega}\triangle},\\
    & & \frac{\partial}{\partial x_{j}} =
    {\rm e}^{\frac{1}{4\omega}\triangle}
    {\rm e}^{\frac{1}{2}\omega\vecvar{x}^{2}}a_{j}
    {\rm e}^{-\frac{1}{2}\omega\vecvar{x}^{2}}
    {\rm e}^{-\frac{1}{4\omega}\triangle},\\
    & & x_{j}\frac{\partial}{\partial x_{j}}=
    {\rm e}^{\frac{1}{4\omega}\triangle}
    {\rm e}^{\frac{1}{2}\omega\vecvar{x}^{2}}n_{j}
    {\rm e}^{-\frac{1}{2}\omega\vecvar{x}^{2}}
    {\rm e}^{-\frac{1}{4\omega}\triangle}.
  \end{eqnarray}
  \label{eqn:Sogo_representation}
\end{subequations}
Then the similarity transformation of the number state is expressed by
the monomial acted by the exponentiation of the Lassalle operators,
\begin{equation}
  T|n_{1},\cdots,n_{N}\rangle
  ={\rm e}^{-\frac{1}{4\omega}{\cal O}_{\rm L}}x_{1}^{n_{1}}
  x_{2}^{n_{2}}\cdots x_{N}^{n_{N}},
\end{equation}
for both $A_{N-1}$ and $B_{N}$ cases. However, as we can see in
\ref{sec:singularity}, acting the Lassalle operator infinitely many times
on a monomial generates essential singularities at $x_{i}=x_{j}$ for
both cases and in addition at $x_{i}=-x_{j}$ and $x_{i}=0$
for the $B_{N}$ case.
Thus we have to consider some escape from such essential singularities
in order to make physical eigenfunctions for the Calogero models.

The keys to such an escape are symmetrization for both cases
and additional restriction to even parity for the $B_{N}$ case.
Here we introduce two symmetrized number states (bosonic states) 
which respectively correspond to the $A_{N-1}$ and $B_{N}$ cases as
\begin{subequations}
  \label{sym-number-state}
  \begin{equation}
    \label{sym-A-number-state}
    |\lambda\rangle
    \Define\sum_{\stackrel{\scriptstyle \sigma\in S_{N}}{\rm distinct}}
    |\lambda_{\sigma(1)},\cdots,\lambda_{\sigma(N)}\rangle
    =m_{\lambda}(\vecvar{a}^{\dagger})|0\rangle,
  \end{equation} 
  \begin{equation}
    \label{sym-B-number-state}
    |2\lambda\rangle
    \Define\sum_{\stackrel{\scriptstyle \sigma\in S_{N}}{\rm distinct}}
    |2\lambda_{\sigma(1)},\cdots,2\lambda_{\sigma(N)}\rangle
    =m_{\lambda}((\vecvar{a}^{\dagger})^{2})|0\rangle
    =m_{2\lambda}(\vecvar{a}^{\dagger})|0\rangle,
  \end{equation}
\end{subequations}
where $\lambda$ and $m_{\lambda}$ are the Young diagram, or a partition 
of a nonnegative integer $|\lambda|\Define\sum_{j=1}^{N}\lambda_{j}$
of at most $N$ parts, and the monomial symmetric function,
\begin{eqnarray}
  \lambda\Define
  \bigl\{\lambda_{1}\ge\lambda_{2}\ge\cdots\ge\lambda_{N}\ge0\bigr\}, \\
  \quad \quad \lambda_{k}, \quad k=1,2,\cdots,N,\mbox{ are integers},
  \nonumber \\
  \label{mono-sym-func}
  m_{\lambda}(\vecvar{x})
  \Define\sum_{\stackrel{\scriptstyle \sigma\in S_{N}}{\rm distinct}}
  x_{1}^{\lambda_{\sigma(1)}}x_{2}^{\lambda_{\sigma(2)}}\cdots
  x_{N}^{\lambda_{\sigma(N)}}.
\end{eqnarray}
Note that the summation over distinct permutations is done so that each
monomial appears only once. A simplified but ambiguous notation 
such as $m_{\lambda}(\vecvar{x^{2}})\Define
m_{\lambda}(x_{1}^{2},\cdots,x_{N}^{2})$ has been introduced
to compactify arguments of multivariable functions.
These symmetrized number states are the simultaneous eigenfunctions
for any symmetrized functions, say, the power sums
$P_{l}(n_{1},\cdots,n_{N})$, of the number operators, 
\begin{subequations}
\begin{eqnarray}
  \label{power-sums}
  P_{l}(\vecvar{n})|\lambda\rangle
  = P_{l}(\lambda)|\lambda\rangle,\\
  P_{l}(\vecvar{n})|2\lambda\rangle
  =P_{l}(2\lambda)|2\lambda\rangle,
\end{eqnarray}
\end{subequations}
where 
\begin{equation}
  P_{l}(\vecvar{x})\Define\sum_{j=1}^{N}(x_{j})^{l}.
\end{equation}
We take the power sums of the number operators,
$P_{l}(\vecvar{n})$, $l=1,2,\cdots,N$, as the set of
commuting conserved operators of the harmonic oscillators.
Consequently we regard the symmetrized number states
(\ref{sym-number-state}) as the states that are uniquely
identified by the set of quantum numbers,
\begin{subequations}
  \begin{eqnarray}
    \bigl\{P_{1}(\lambda),\cdots,P_{N}(\lambda)\bigr\},\\
    \bigl\{P_{1}(2\lambda),\cdots,P_{N}(2\lambda)\bigr\},
  \end{eqnarray}
\end{subequations}
for the $A_{N-1}$ and $B_{N}$ cases, respectively. Since they are 
eigenfunctions of Hermitian operators
without degeneracy, they form orthogonal bases for the harmonic
oscillators.

We define the dual bases for the states \myref{sym-A-number-state}
and \myref{sym-B-number-state} by
\begin{subequations}
\begin{eqnarray}
 \langle\lambda|\Define\langle 0|m_{\lambda}(\vecvar{a}),\\
 \langle 2\lambda|\Define\langle 0|m_{2\lambda}(\vecvar{a}),
\end{eqnarray}
\end{subequations}
where the vacuum bra $\langle 0|=e^{-\frac{1}{2}\vecvar{x}^{2}}$.
Rewriting the Young diagram $\lambda$ as
\begin{equation}
 \lambda
 =\{\underbrace{\lambda_{1},\cdots,\lambda_{1}}_{n_{1}},
    \underbrace{\lambda_{n_{1}+1},\cdots,\lambda_{n_{1}+1}}_{n_{2}},
    \lambda_{n_{1}+n_{2}+1},\cdots\}
 =\{\lambda_{1},\cdots,\lambda_{N}\},
\end{equation}
we confirm the orthogonality of the symmetrized number states,
\begin{subequations}
\begin{eqnarray}
 \langle \mu|\lambda\rangle
 =\frac{N!}{n_{1}!n_{2}!\cdots}\prod_{j=1}^{N}\lambda_{j}!\,
  \langle 0|0\rangle\,\delta_{\lambda\mu},\\
 \langle 2\mu|2\lambda\rangle
 =\frac{N!}{n_{1}!n_{2}!\cdots}\prod_{j=1}^{N}(2\lambda_{j})!\,
  \langle 0|0\rangle\,\delta_{\lambda\mu}.
\label{orthogonality}
\end{eqnarray}
\end{subequations}
We note that the dual bases are Hermitian conjugates of the symmetrized
number states, which reflects the fact that the number operators of the
harmonic oscillators are Hermitian operators.

By the transformation of symmetrized number states,
\begin{subequations}
\begin{eqnarray}
 T^{(A)}|\lambda\rangle
 =e^{-\frac{1}{4\omega}{\cal O}_{\rm L}^{(A)}}m_{\lambda}(\vecvar{x})
 \Define M_{\lambda}(\vecvar{x};1/a,\omega),\\
 T^{(B)}|2\lambda\rangle
 =e^{-\frac{1}{4\omega}{\cal O}_{\rm L}^{(B)}}m_{2\lambda}(\vecvar{x})
 \Define Y_{\lambda}(\vecvar{x};1/a,1/b,\omega),
\end{eqnarray}
\end{subequations}
we get the eigenfunctions of the Calogero models \myref{A-calogero2}
and \myref{B-calogero2}, or the symmetric polynomial parts of the
eigenfunctions of the original Calogero models \myref{A-calogero1}
and \myref{B-calogero1}. They are, indeed, symmetric polynomials
and do not have any essential singularities. Detailed discussions on
the cancellation of essential singularities are presented in 
\ref{sec:singularity}.

Now we transform the symmetrized number state and get
orthogonal bases for the Calogero models 
\myref{A-calogero1} and \myref{B-calogero1}.
We introduce the creation and annihilation operators for the Calogero 
model as
\begin{subequations}
\begin{eqnarray}
  b_{j}^{+}
  \Define T a_{j}^{\dagger}T^{-1}
  =\e^{-\frac{1}{4\omega}{\cal O}_{{\rm L}}}x_{j}
   \e^{\frac{1}{4\omega}{\cal O}_{{\rm L}}}, \\
  b_{j}
  \Define T a_{j}T^{-1}
  =\e^{-\frac{1}{4\omega}{\cal O}_{{\rm L}}}
   \frac{\partial}{\partial x_{j}}
   \e^{\frac{1}{4\omega}{\cal O}_{{\rm L}}}, \\
  \nu_{j}\Define b_{j}^{+} b_{j}.
\end{eqnarray}
\label{b-ope}
\end{subequations}
Including the action to the ground state wave function, we obtain
the creation-annihilation operators for the original Calogero models
\myref{A-calogero1} and \myref{B-calogero1},
\begin{subequations}
\begin{eqnarray}
 \hat{b}_{j}^{+}\Define\Psi_{\rm g}b_{j}^{+}(\Psi_{\rm g})^{-1},\\
 \hat{b}_{j}\Define\Psi_{\rm g}b_{j}(\Psi_{\rm g})^{-1},\\
 \hat{\nu}_{j}\Define\hat{b}_{j}^{+}\hat{b}_{j}.
\end{eqnarray}
\end{subequations}
In terms of the above creation operators,
the eigenfunctions of the original Calogero models are
\begin{subequations}
  \begin{eqnarray}
    \fl |\lambda\rangle^{(A)}
    \Define\Psi_{\rm g}^{(A)}T^{(A)}|\lambda\rangle
    =\Psi_{\rm g}^{(A)}\e^{-\frac{1}{4\omega}{\cal O}_{{\rm L}}^{(A)}}
    m_{\lambda}(\mbox{\boldmath $x$})
    =m_{\lambda}(\hat{\vecvar{b}}^{(A)+})|0\rangle^{(A)}, 
    \label{A-C-eigenfunc}\\
    \fl |\lambda\rangle^{(B)}
    \Define\Psi_{\rm g}^{(B)}T^{(B)}|2\lambda\rangle
    =\Psi_{\rm g}^{(B)}\e^{-\frac{1}{4\omega}{\cal O}_{{\rm L}}^{(B)}}
    m_{\lambda}(\vecvar{x}^{2})
    =m_{\lambda}((\hat{\vecvar{b}}^{(B)+})^{2})|0\rangle^{(B)},
    \label{B-C-eigenfunc}
  \end{eqnarray}
  \label{last-eigenfunc}
\end{subequations}
where $|0\rangle^{(A)}\Define\Psi_{\rm g}^{(A)}$ and
$|0\rangle^{(B)}\Define\Psi_{\rm g}^{(B)}$ are the ground states for the
original Calogero models. The eigenfunctions \myref{last-eigenfunc}
simultaneously diagonalize all the mutually commuting conserved operators,
\begin{equation}
 \label{com-cons-op}
 P_{l}(\hat{\vecvar{\nu}})=\sum_{j=1}^{N}(\hat{\nu}_{j})^{l},
 \quad l=1,2,\cdots,N.
\end{equation}
The dual bases are defined in a similar way to that of the quantum
harmonic oscillators,
\begin{subequations}
\begin{eqnarray}
 ^{(A)}\langle\lambda|
 \Define \,^{(A)}\langle 0|m_{\lambda}(\hat{\vecvar{b}}^{(A)}),\\
 ^{(B)}\langle\lambda|
 \Define \,^{(B)}\langle 0|m_{\lambda}((\hat{\vecvar{b}}^{(A)})^{2}),
\end{eqnarray}
\end{subequations}
where  $ ^{(A)}\langle 0|=\Psi_{\rm g}^{(A)}$
and $ ^{(B)}\langle 0|=\Psi_{\rm g}^{(B)}$.
Their orthogonality is also confirmed in a similar way,
\begin{subequations}
\begin{eqnarray}
 ^{(A)}\langle\mu|\lambda\rangle^{(A)}
 =\frac{N!}{n_{1}!n_{2}!\cdots}\prod_{j=1}^{N}\lambda_{j}!
  \, ^{(A)}\langle 0|0\rangle^{(A)}\,\delta_{\lambda\mu},\\
 ^{(B)}\langle\mu|\lambda\rangle^{(B)}
 =\frac{N!}{n_{1}!n_{2}!\cdots}\prod_{j=1}^{N}(2\lambda_{j})!
  \, ^{(B)}\langle 0|0\rangle^{(B)}\,\delta_{\lambda\mu},
\end{eqnarray}
 \label{cal-orthogonality}
\end{subequations}
and the vacuum normalizatin constants are
\begin{subequations}
\begin{eqnarray}
 \fl ^{(A)}\langle 0|0\rangle^{(A)}
 =(\frac{1}{2\omega})^{\frac{N(Na+(1-a))}{2}}(2\pi)^{\frac{N}{2}}N!
  \nonumber\\
  \times\prod_{1\le j<k\le N}
  \frac{\Gamma((k-j+1)a)\Gamma(1+(k-j+1)a)}{\Gamma((k-j)a)\Gamma(1+(k-j)a)}
  \nonumber\\
  \times\prod_{1\le j\le N}\Gamma(1+(N-j)a),\\
 \fl ^{(B)}\langle 0|0\rangle^{(B)}
 =(\frac{1}{\omega})^{N(N-1)a+N(b+\frac{1}{2})}N!
  \nonumber\\
  \times\prod_{1\le j<k\le N}
  \frac{\Gamma((k-j+1)a)\Gamma(1+(k-j+1)a)}{\Gamma((k-j)a)\Gamma(1+(k-j)a)}
  \nonumber\\
  \times\prod_{1\le j\le N}\Gamma(1+(N-j)a)\Gamma((N-j)a+b+\frac{1}{2}),
\end{eqnarray}
\end{subequations}
where $\Gamma(z)$ denotes the gamma functions.
A proof of the vacuum normalization constants is given 
in~\cite{Baker,vanDiejen}.

As is similar to the triangularity of the Hi-Jack polynomials%
~\cite{Ujino4}, polynomial parts of 
these eigenfunctions possess the triangularity,
\begin{subequations}
  \begin{eqnarray}
    & & \fl M_{\lambda}(\mbox{\boldmath $x$};1/a,\omega)
    = m_{\lambda}(\vecvar{x})+\sum_{
    \stackrel{\scriptstyle \mu \stackrel{\rm d}{<}\lambda,\;|\mu|<|\lambda|}
    {|\mu|\equiv|\lambda| \! \pmod{2}}}
    \left(-\frac{1}{4\omega}\right)^{(|\lambda|-|\mu|)/2}\!\!\!
    w^{(A)}_{\lambda\mu}(a)m_{\mu}(\mbox{\boldmath $x$}),\\
    & & \fl Y_{\lambda}(\vecvar{x};1/a,1/b,\omega)
    = m_{\lambda}(\vecvar{x}^{2})
    +\sum_{\mu \stackrel{\rm d}{<}\lambda,\;|\mu|<|\lambda|}
    \left(-\frac{1}{4\omega}\right)^{|\lambda|-|\mu|}\!\!\!
    w^{(B)}_{\lambda\mu}(a,b)m_{\mu}(\vecvar{x}^{2}),
  \end{eqnarray}
  \label{sym-orth-basis}
\end{subequations}
with respect to the weak dominance order,
\begin{equation}
  \label{dominance-order}
  \mu \stackrel{\rm d}{<}\lambda 
  \Leftrightarrow 
  \mu\neq\lambda\mbox{ and }
  \sum_{k=1}^{l}\mu_{k}\leq\sum_{k=1}^{l}\lambda_{k} \;
  \mbox{ for all } l=1,2,\cdots,N.
\end{equation}
The coefficients
$w^{(A)}_{\lambda\mu}(a)$ and $w^{(B)}_{\lambda\mu}(a,b)$ are polynomials
of the coupling parameters with integer coefficients,
which is similar to integrality of the Jack and Hi-Jack
polynomials~\cite{Macdonald,Ujino4}.
Explicit forms of some of the above symmetric orthogonal
polynomials are shown in \ref{sec:explicit_forms}. We note that 
the above orthogonal symmetric polynomials, \myref{A-C-eigenfunc}
and \myref{B-C-eigenfunc},
can be interpreted as a
multivariable generalization of the Hermite polynomial and
a multivariable generalization of the Laguerre polynomial, respectively.


\section{Relationships between new and known orthogonal bases}
\label{sec:relation}
As we mentioned before, there are known orthogonal bases for
the $A_{N-1}$- and $B_{N}$-Calogero models, namely, the Hi-Jack
and the multivariable Laguerre polynomials that are variants of the
Jack polynomials. We shall compare the new and the known orthogonal
bases here.

We use the formulae that relate 
the Hi-Jack polynomial $j_{\lambda}(\vecvar{x};1/a,\omega)$
and the multivariable Laguerre polynomial
$l_{\lambda}(\vecvar{x};1/a,1/b,\omega)$ with
the Jack polynomial $J_{\lambda}(\vecvar{x};1/a)$%
~\cite{Baker,Sogo,Lassalle1,Lassalle2},
\begin{subequations}
  \label{Baker}
  \begin{eqnarray}
    j_{\lambda}(\vecvar{x};1/a,\omega)
    =J_{\lambda}(\vecvar{\alpha}^{(A)\dagger};1/a)\cdot 1
    =\e^{-\frac{1}{4\omega}{\cal O}_{{\rm L}}^{(A)}} 
    J_{\lambda}(\vecvar{x};1/a), \\
    l_{\lambda}(\vecvar{x};1/a,1/b,\omega)
    =J_{\lambda}((\vecvar{\alpha}^{(B)\dagger})^{2};1/a)\cdot 1
    =\e^{-\frac{1}{4\omega}{\cal O}_{{\rm L}}^{(B)}} 
    J_{\lambda}(\vecvar{x}^{2};1/a).
  \end{eqnarray}
\end{subequations}
The definitions of the Dunkl operators, $\alpha_{k}^{(A)\dagger}$ and
$\alpha_{k}^{(B)\dagger}$, and the Jack polynomial
$J_{\lambda}(\mbox{\boldmath $x$})$ are presented in
\ref{sec:Dunkl-f}. Those variants of the Jack polynomial are
respectively different from the new orthogonal bases
(\ref{last-eigenfunc}) for the corresponding models obtained 
in the previous section and do not diagonalize the conserved 
operators (\ref{com-cons-op}). On the other hand, the Hi-Jack and
the multivariable Laguerre polynomials are uniquely identified as
the simultaneous eigenfunctions of corresponding sets of conserved
operators, $I_{k}^{(A)}$ and $I_{k}^{(B)}$ for $k=1,2,\cdots,N$,
given by the Dunkl operator formulation. This means that the new orthogonal
bases are not the simultaneous eigenfunctions for these conserved
operators.

For the sake of fairness , we should note that the ``new''
orthogonal bases are, in a sense, ``old'' because they are nothing
but what was given by Brink, Hansson, Konstein and Vasiliev
for the $A_{N-1}$-Calogero model~\cite{Brink1,Brink2}. 
A proof is as follows.
The Jack polynomials have triangular expansion in
the monomial symmetric functions,
\begin{equation}
  \label{linear-relation}
  J_{\lambda}(\mbox{\boldmath $x$};1/a)
  =\sum_{\mu\stackrel{\rm D}{\leq} \lambda}
   v_{\lambda\mu}(a)m_{\mu}(\mbox{\boldmath $x$})
  , \quad v_{\lambda\lambda}=1,
\end{equation}
with respect to the dominance order,
\begin{equation}
  \mu\stackrel{\rm D}{\leq}\lambda \Leftrightarrow
  |\mu|=|\lambda|\mbox{ and }\sum_{j=1}^{l}\mu_{j}\leq
  \sum_{j=1}^{l}\lambda_{j}\mbox{ for all }l=1,2,\cdots,N.
\end{equation}
Since the triangular matrix $v_{\lambda\mu}(a)$ has its inverse,
we have
\begin{equation}
  \label{inverse}
  m_{\lambda}(\mbox{\boldmath $x$})
  =\sum_{\mu\stackrel{D}{\leq} \lambda}
   (v^{-1})_{\lambda\mu}(a)J_{\mu}(\mbox{\boldmath $x$};1/a).
\end{equation}
Applying the above transformation to the formulae \myref{Baker},
we have
\begin{subequations}
\begin{eqnarray}
  m_{\lambda}
  (\vecvar{\alpha}^{(A)\dagger})
  \cdot 1
  & = &\sum_{\mu \stackrel{D}{\leq}\lambda }
  (v^{-1})_{\lambda\mu}(a)
  J_{\mu}(\vecvar{\alpha}^{(A) \dagger};1/a)\cdot 1 \nonumber\\
  & = & \e^{-\frac{1}{4\omega}{\cal O}_{{\rm L}}^{(A)}}
  \sum_{\mu \stackrel{D}{\leq}\lambda }(v^{-1})_{\lambda\mu}(a)
  J_{\mu}(\vecvar{x};1/a)\cdot 1 \nonumber\\
  & = & \e^{-\frac{1}{4\omega}{\cal O}_{{\rm L}}^{(A)}}
  m_{\lambda}(\mbox{\boldmath $x$})
  =m_{\lambda}(\vecvar{b}^{(A)+})\cdot 1,\\
  m_{\lambda}((\vecvar{\alpha}^{(B)\dagger})^{2})\cdot 1
  & = & \sum_{\mu \stackrel{D}{\leq}\lambda }
   (v^{-1})_{\lambda\mu}(a)
   J_{\mu}((\vecvar{\alpha}^{(B)\dagger})^{2};1/a)\cdot 1\nonumber\\
  & = & \e^{-\frac{1}{4\omega}{\cal O}_{{\rm L}}^{(B)}}
   \sum_{\mu \stackrel{D}{\leq}\lambda }(v^{-1})_{\lambda\mu}(a)
   J_{\mu}(\vecvar{x}^{2};1/a)\cdot 1\nonumber\\
  & = & \e^{-\frac{1}{4\omega}{\cal O}_{{\rm L}}^{(B)}}
   m_{\lambda}(\vecvar{x}^{2})
  =m_{\lambda}((\vecvar{b}^{(B)+})^{2})\cdot 1,
\end{eqnarray}
\end{subequations}
which show the ``new'' orthogonal basis for the $A_{N-1}$-Calogero model
is nothing but the basis given in \cite{Brink1,Brink2},
though its orthogonality and corresponding conserved operators were
not given.
We note that the creation-annihilation operators,
$b_{j}^{+}$ and $b_{j}$, cannot be the same as the Dunkl operators,
$\alpha_{j}^{\dagger}$ and  $\alpha_{j}$, respectively.
If so, then the two sets of conserved operators $P_{k}(\vecvar{\nu})$
and $I_{k}$ become the same and the corresponding simultaneous eigenfunctions
also must be the same, which is contradictory. We can also directly verify it
by calculating the forms of the creation-annihilation operators.

Since the transition matrix $v_{\lambda\mu}$ that relates the new and
the known orthogonal bases is not a unitary but triangular matrix,
it seems rather strange at first sight that the new orthogonal basis
is indeed an orthogonal basis. This strange observation comes from
the fact that the new sets of conserved operators 
$P_{l}(\hat{\vecvar{\nu}})$ are not Hermitian, but self-dual with respect
to the exchange of creation-annihilation operators, 
$\hat{b}_{l}^{+}\leftrightarrow\hat{b}_{l}$. That is the reason why the
new orthogonal bases are orthogonal with respect to the inner product
\myref{cal-orthogonality}. On the other hand, the conserved operators given
by the Dunkl operator formulation including the action to the ground
state wave function, $\hat{I}_{k}\Define\Psi_{\rm g}I_{k}\Psi_{\rm g}^{-1}$
are Hermitian operators, $\hat{I}_{k}^{\dagger}=\hat{I}_{k}$. That
explains why the Hi-Jack and the multi-variable Laguerre polynomials are
orthogonal with respect to the conventional Hermitian inner product,
\begin{subequations}
\begin{eqnarray}
 \int_{-\infty}^{\infty}\!\!\cdots\int_{-\infty}^{\infty}
 \prod_{j=1}^{N}\d x_{j}
 |\Psi_{\rm g}^{(A)}(\vecvar{x})|^{2}\,\,
 j_{\lambda}^{\dagger}(\vecvar{x})j_{\mu}(\vecvar{x})
 \propto\delta_{\lambda\mu},
 \label{D-A-inprod} \\
 \int_{-\infty}^{\infty}\!\!\cdots\int_{-\infty}^{\infty}
 \prod_{j=1}^{N}\d x_{j}
 |\Psi_{\rm g}^{(B)}(\vecvar{x})|^{2}\,\,
 l_{\lambda}^{\dagger}(\vecvar{x})l_{\mu}(\vecvar{x})
 \propto\delta_{\lambda\mu}.
 \label{D-B-inprod} 
\end{eqnarray}
\end{subequations}
In our normalization, the above polynomials are real functions, 
$j_{\lambda}^{\dagger}(\vecvar{x})=j_{\lambda}(\vecvar{x})$,
$l_{\lambda}^{\dagger}(\vecvar{x})=l_{\lambda}(\vecvar{x})$.
Comparing the two different orthogonal bases and inner products, we notice
that the dual bases $ ^{(A)}\langle\lambda|$ and $ ^{(B)}\langle\lambda|$
are identified with the ``rotation'' of the variants of the Jack polynomial
up to normalization,
\begin{subequations}
\begin{eqnarray}
 ^{(A)}\langle\lambda|
 &\propto&\Psi_{\rm g}^{(A)}(\vecvar{x})\sum_{\mu}
  v_{\lambda\mu}(a)j_{\mu}(\vecvar{x}) \nonumber\\
 &=&\Psi_{\rm g}^{(A)}(\vecvar{x})\sum_{\mu\rho}
  v_{\lambda\mu}(a)v_{\mu\rho}(a)M_{\rho}(\vecvar{x}),\\
 ^{(B)}\langle\lambda|
 &\propto&\Psi_{\rm g}^{(B)}(\vecvar{x})\sum_{\mu}
  v_{\lambda\mu}(a)l_{\mu}(\vecvar{x}) \nonumber\\
 &=&\Psi_{\rm g}^{(B)}(\vecvar{x})\sum_{\mu\rho}
  v_{\lambda\mu}(a)v_{\mu\rho}(a)Y_{\rho}(\vecvar{x}).
\end{eqnarray}
\end{subequations}
The above identification of the dual bases is, at least, valid in the
consideration of the inner product. Thus both new and known orthogonal
bases give the same thermodynamics quantities calculated by the trace
formula. However, we should note that a naive identification of the 
dual bases as functions themselves has difficulties. Further
considerations on this point are left for future studies.

From the discussions above, we conclude that each Calogero model has
at least two sets of commuting conserved operators which are algebraically
inequivalent to each other. We also conclude that two conserved operators
respectively picked up from two different sets
do not commute $[P_{n}(\hat{\vecvar{\nu}}),\hat{I}_{k}]\neq 0$, 
for $n,k\neq 1$.
Its Hilbert space also has two different
orthogonal bases with respect to two ``different'' inner products 
that respectively correspond to the simultaneous
eigenfunctions of two sets of commuting conserved operators.
This peculiar fact may be due to the large degeneracy of the
eigenvalue of the Calogero models,
\begin{subequations}
  \begin{eqnarray}
    H_{\rm C}^{(A)}M_{\lambda}(\vecvar{x};1/a,\omega)
    =\omega|\lambda|M_{\lambda}(\vecvar{x};1/a,\omega),\\
    H_{\rm C}^{(B)}Y_{\lambda}(\vecvar{x};1/a,1/b,\omega)
    =2\omega|\lambda|Y_{\lambda}(\vecvar{x};1/a,1/b,\omega).
  \end{eqnarray}
\end{subequations}
For a particular eigenvalue, say $\omega n$ and $2\omega n$,
the degeneracy is given by
the number of the Young diagrams $\lambda$ such that $|\lambda|=n$,
namely, the number of partitions not exceeding $N$ parts.

The above peculiar story reminds us of another example of such a
story, i.e.,  the wave functions of the hydrogen atom~\cite{Schiff_1}.
The well-known eigenfunctions for the hydrogen atom,
\begin{equation}
  H_{\rm H}\Phi(\vecvar{r})\Define\Bigl(
  \frac{1}{2m}\vecvar{p}^{2}
  -\frac{k}{r}\Bigr)\Phi(\vecvar{r})=E\Phi(\vecvar{r}),
\end{equation}
where $\vecvar{p}$,
$\vecvar{r}$ and $r$ denote the momentum operator, the coordinate vector
and its norm in three dimensional space,
are obtained by separation of variables in spherical coordinates.
They are the simultaneous eigenfunctions of the Hamiltonian $H_{\rm H}$,
the total
angular momentum $\vecvar{L}^{2}$ and its $z$-axis component $L_{z}$
where $\vecvar{L}=\vecvar{r}\times\vecvar{p}$, and
are expressed by the product
of the spherical harmonics and the associated Laguerre polynomial.
On the other hand, the Schr\"{o}dinger equation can be solved by
separation of variables in parabolic coordinates and results in the
wave functions that contain a product of two associated
Laguerre polynomials.
They simultaneously diagonalize $H_{\rm H}$, $L_{z}$
and the $z$-axis component of the Runge-Lenz-Pauli vector $M_{z}$,
\begin{equation}
  \vecvar{M}\Define\frac{1}{2m}(\vecvar{p}\times\vecvar{L}-
  \vecvar{L}\times\vecvar{p})-\frac{k}{r}\vecvar{r}.
\end{equation}
This is another conserved operator of the hydrogen atom.
The total angular momentum and the Runge-Lenz-Pauli
vector are algebraically different because of the relation,
$[M_{3},\vecvar{L}^{2}]\neq 0$.
The two solutions respectively derived by two ways are different,
and the former
is a linear combination of the latter and vice versa.
This is quite similar to what  we have
observed for the Calogero models.
Behind the story for the hydrogen atom, there is the $O(4)$ dynamical
symmetry. We stress that the quest for some hidden dynamical symmetry
for the Calogero model must be an interesting future problem.

A similar transformation for the Sutherland model into a
decoupled system has not been found so far. This seems rather strange
at first because the commuting
conserved operators and the Dunkl operators of the 
Calogero and Sutherland models share the same algebraic structure,
and become exactly the same in the limit,
$\omega\rightarrow\infty$~\cite{Ujino3,Ujino4,Polychronakos}.
The difference of the two models is the structure of the Hamiltonian.
While the Calogero Hamiltonian is the simplest
conserved operator $I_{1}$, the Sutherland Hamiltonian corresponds
to the second conserved operator $I_{2}$. We have proved that the
second conserved operator $I_{2}$ can not be constructed from the
power sums of the number operators $P_{n}(\vecvar{\nu})$.
We think that the point causes the critical
difficulty in the application of such a similarity transformation method
to the Sutherland model.


\section{Summary}\label{sec:summary}
We have studied an algebraic construction of new orthogonal bases for the
$A_{N-1}$- and $B_{N}$-Calogero models and their conserved operators
by means of similarity transformations to decoupled quantum harmonic
oscillators. Our idea is just pulling the number operators and number
states back to the Hilbert space of the Calogero models by the similarity
transformations. We have pointed out that 
a delicate property of the exponentiation of the Lassalle operators
which yields essential singularities requires its operands to be
symmetric functions for both models.
For the case of the $B_{N}$-Calogero model, the property
further demands its operands to be even functions.
Thus we introduce the symmetrized number state, which is nothing but
the bosonic state for the quantum harmonic oscillators, and restrict
its parity to be even for the $B_{N}$ case. The symmetrized number
state is uniquely determined as the symmetric simultaneous eigenfunctions
for the power sums of the number operators, $P_{l}(\vecvar{n})$,
$l=1,2,\cdots,N$, and spans the orthogonal basis of the $N$ decoupled
quantum harmonic oscillators. Since the conserved operators 
$P_{l}(\hat{\vecvar{\nu}})$, $l=2,\cdots,N$, are not Hermitian,
the definition of the inner products for the new orthogonal bases is
different from the conventional Hermitian inner product.
Consequently we have obtained two sets of
commuting conserved operators and two sets of orthogonal symmetric polynomials
as the simultaneous eigenfunctions of the conserved operators.
The orthogonal symmetric polynomials span new orthogonal bases for the
$A_{N-1}$- and $B_{N}$-Calogero models.

The two Calogero models have known orthogonal bases that are spanned by
the Hi-Jack and the multivariable Laguerre polynomials, which are
uniquely identified as the simultaneous eigenfunctions of the conserved
operators constructed by the Dunkl operator formulations for the models.
Comparison of the new and the known orthogonal bases reveals that they are
different, though both are considered to be multivariable generalizations
of the Hermite polynomials and those of the Laguerre polynomials.
The fact means that the conserved operators given by the similarity
transformations and those given by the Dunkl operator formulations
do not commute. Thus the $A_{N-1}$- and $B_{N}$-Calogero models respectively
have two sets of commuting conserved operators that are algebraically
inequivalent to each other and two orthogonal bases that respectively
correspond to the simultaneous eigenfunctions for each set of commuting
conserved operators. We have conjectured that this peculiar fact implies
some hidden dynamical symmetry for the models, as is the case with the
hydrogen atom. We have shown triangularity and integrality that appear
in the expansion form of the new orthogonal symmetric polynomials
with respect to the monomial symmetric functions. For the $A_{N-1}$ case,
the ``new'' orthogonal basis turns out to be the same basis that was given
by Brink, Hansson, Konstein and Vasiliev.
Still we stress it is new as an orthogonal
basis because its orthogonality and corresponding conserved operators are
presented in this paper.
We have discussed the difficulty in the application of the similarity
transformation method to the Sutherland model from the algebraic
structural point of view.

In short, we have completed a construction of the symmetric Fock space of
the Calogero models. Recently, an
extension of the similarity transformation method to the non-symmetric case,
which gives complete Fock space for the Calogero Hamiltonians,
was announced~\cite{Ujino7}.
We expect that these Fock spaces will be a useful tool for
calculation of various kinds of quantities such as the Green function 
and correlation functions of the Calogero models.

\section*{Acknowledgements}
One of the authors (HU) appreciates Research Fellowships of the
Japan Society for the Promotion of Science for Young Scientists.


\appendix


\section{Dunkl operator formulation}
 \label{sec:Dunkl-f}
We briefly summarize the Dunkl operator formulations for the
$A_{N-1}$- and $B_{N}$-Calogero and Sutherland models, which are
essential tools to study the algebraic structure, integrability and
variants of the Jack polynomial~\cite{Dunkl,Polychronakos}.

\subsection{$A_{N-1}$-Calogero and Sutherland models}
The Dunkl operators for the $A_{N-1}$-Sutherland model are given by
\begin{eqnarray}
  \nabla_{l}^{(A)}=\frac{\partial}{\partial x_{l}}
                   +a\sum_{\stackrel{\scriptstyle k=1}{k\neq l}}^{N}
                   \frac{1}{x_{l}-x_{k}}(1-K_{lk}),\nonumber\\
  x_{l},\nonumber\\
  D_{l}^{(A)}=x_{l}\nabla _{l}^{(A)},
  \label{A-S-Dunkl}
\end{eqnarray}
where $K_{lk}$ is the coordinate exchange operator that is defined by
the action on multi-variable functions of
$\mbox{\boldmath $x$}=(x_{1},\cdots,x_{N})$,
\begin{equation}
  \label{exchange}
  (K_{lk}f)(\cdots,x_{l},\cdots,x_{k},\cdots)
  =f(\cdots,x_{k},\cdots,x_{l},\cdots).
\end{equation}
Commutation relations among the Dunkl operators are given by
\begin{eqnarray}
  \bigl[\nabla_{l}^{(A)},\nabla_{m}^{(A)}\bigr]=0, 
  \quad \quad \bigl[x_{l},x_{m}\bigr]=0,\nonumber\\
  \bigl[\nabla_{l}^{(A)},x_{m}\bigr]
  =\delta_{lm}(1+a\sum_{\stackrel{\scriptstyle k=1}{k\neq l}}^{N} K_{lk})
  -a(1-\delta_{lm})K_{lm},\nonumber\\
  \bigl[D_{l}^{(A)},D_{m}^{(A)}\bigr]=a(D_{m}^{(A)}-D_{l}^{(A)})K_{lm},
  \nonumber\\
  \nabla_{l}^{(A)}\cdot 1=0.
  \label{A-S-commutation}
\end{eqnarray}
A set of mutually commuting conserved operators for the $A_{N-1}$-Sutherland 
model are given by 
\begin{equation}
  \label{A-S-conserved}
  L_{k}^{(A)}=\sum_{l=1}^{N}(D_{l}^{(A)})^{k}\Bigr|_{\rm Sym},\quad
  \bigl[L_{k}^{(A)},L_{m}^{(A)}\bigr]=0 \quad k,m=1,2,\cdots,N,
\end{equation}
where the symbol $\Bigr|_{\rm Sym}$ means that the action of the operator is 
restricted to symmetric functions, 
$K_{lk}f(\mbox{\boldmath $x$})=f(\mbox{\boldmath $x$})$. 
The Hamiltonian of the 
$A_{N-1}$-Sutherland model corresponds to the second conserved 
operator $L_{2}^{(A)}$. In terms of the $A_{N-1}$-Sutherland Hamiltonian,
the Jack symmetric polynomials
$J_{\lambda}(\vecvar{x};1/a)$ are uniquely defined by~\cite{Macdonald}
\begin{eqnarray}
  & & L_{2}^{(A)}J_{\lambda}(\vecvar{x};1/a)
  = \sum_{k=1}^{N}
  \bigl(\lambda_{k}^{2}+a(N+1-2k)\lambda_{k}\bigr)
  J_{\lambda}(\vecvar{x};1/a),\nonumber\\
  & & J_{\lambda}(\vecvar{x};1/a) = 
  \sum_{\mu\ledo\lambda}v_{\lambda\mu}(a)m_{\mu}(\vecvar{x}),\quad
  v_{\lambda\lambda}(a) = 1,
  \label{eqn:Jack_definition}
\end{eqnarray}
where $\lambda$ and $\mu$ are the Young diagrams.
The symbol $\ledo$ is the dominance order among the Young
diagrams \cite{Macdonald},
\begin{equation}
  \mu\ledo\lambda\Leftrightarrow\sum_{k=1}^{N}
  \mu_{k}=\sum_{k=1}^{N}\lambda_{k}
  \mbox{ and }\sum_{k=1}^{l}\mu_{k}\leq\sum_{k=1}^{l}\lambda_{k}
  \mbox{ for all } l.
  \label{eqn:Dominance_ordering}
\end{equation}
Note that the dominance order is not a total order but a partial order.
Since the Jack symmetric polynomials 
diagonalize all the mutually commuting conserved operators $L_{k}^{(A)}$ 
simultaneously, they form the orthogonal basis of the $A_{N-1}$-Sutherland
model~\cite{Sutherland1,Sutherland2,Jack,Macdonald,Stanley,Forrester1,Forrester2,Ha1,Ha2,Lapointe1,Lapointe2}.

A similar formulation is also applicable 
to the $A_{N-1}$-Calogero model. The Dunkl operators for the 
$A_{N-1}$-Calogero model are
\begin{eqnarray}
  &&\alpha_{l}^{(A)}=\nabla_{l}^{(A)}, \nonumber\\
  &&\alpha_{l}^{(A)\dagger}=-\frac{1}{2\omega}\nabla_{l}^{(A)}+x_{l},
  \nonumber\\
  &&d_{l}^{(A)}=\alpha_{l}^{(A)\dagger}\alpha_{l}^{(A)}.
  \label{A-C-Dunkl}
\end{eqnarray}
The above Dunkl operators
are a one-parameter deformation of those for the
$A_{N-1}$-Sutherland model and the former reduces to the latter in the limit,
$\omega\rightarrow\infty$.
The Dunkl operators for the $A_{N-1}$-Calogero model satisfy the
commutation relations,
\begin{eqnarray}
  \label{A-C-commutation}
  \bigl[\alpha_{l}^{(A)},\alpha_{m}^{(A)}\bigr]=0, \quad \quad
  \bigl[\alpha_{l}^{(A)\dagger},\alpha_{m}^{(A)\dagger}\bigr]=0,
  \nonumber\\
  \bigl[\alpha_{l}^{(A)},\alpha_{m}^{(A)\dagger}\bigr]
   =\delta_{lm}(1+a\sum_{\stackrel{\scriptstyle k=1}{k\neq l}}^{N}
   K_{lk})-a(1-\delta_{lm})K_{lm}, \nonumber\\
  \bigl[d_{l}^{(A)},d_{m}^{(A)}\bigr]=a(d_{m}^{(A)}-d_{l}^{(A)})K_{lm},
  \nonumber\\
  \alpha_{l}^{(A)}\cdot 1=0,
\end{eqnarray}
which are exactly the same as those for the Sutherland model
\myref{A-S-commutation}. Thus the Dunkl operators for
the $A_{N-1}$-Calogero and Sutherland
models share the same algebraic structure~\cite{Ujino3,Ujino4,Polychronakos}.
Commuting conserved operators for the Calogero model are obtained
in a similar way to \myref{A-S-conserved},
\begin{equation}
  \label{A-C-conserved}
  I_{k}^{(A)}=\sum_{l=1}^{N}(d_{l}^{(A)})^{k}\Bigr|_{\rm Sym},\quad
  \bigl[I_{k}^{(A)},I_{m}^{(A)}\bigr]=0 \quad k,m=1,2,\cdots,N.
\end{equation}
The correspondences between the operators for the two models are
\begin{equation}
  \label{correspondence}
  \alpha_{l}^{(A)}\leftrightarrow\nabla_{l}^{(A)},\quad 
  \alpha_{l}^{(A)\dagger}\leftrightarrow x_{l}, \quad
  d_{l}^{(A)}\leftrightarrow D_{l}^{(A)},\quad
  I_{l}^{(A)}\leftrightarrow L_{l}^{(A)}.
\end{equation}
The commutator algebra of these operators is 
translated by the correspondences in \myref{correspondence}.
The first conserved operator 
$I_{1}^{(A)}$ is identified with the Hamiltonian of the $A_{N-1}$-Calogero 
model, $\omega I_{1}^{(A)}=H_{\rm C}^{(A)}$. Because of these 
correspondences, the Jack symmetric polynomials are transformed into the 
orthogonal basis for the $A_{N-1}$-Calogero model which are called Hi-Jack 
symmetric polynomials, $j_{\lambda}(\vecvar{x})
=J_{\lambda}(\alpha_{1}^{(A)\dagger},\cdots,\alpha_{N}^{(A)\dagger})
\cdot 1$~\cite{Ujino4,Ujino5}.


\subsection{$B_{N}$-Calogero and Sutherland models}
Similar to the $A_{N-1}$ case,
there are Dunkl operator formulations for 
the $B_{N}$-Calogero and Sutherland models.
The Dunkl operators for the $B_{N}$-Sutherland model are
\begin{eqnarray}
  \label{B-S-Dunkl}
  \nabla_{l}^{(B)} & = &
  \frac{\partial}{\partial x_{l}}+\frac{b}{x_{l}}(1-P_{l})
  \nonumber \\
  & & 
  +a\!\!\sum_{\stackrel{\scriptstyle k=1}{k\neq l}}^{N}
  \!\!\left(\frac{1}{x_{l}-x_{k}}(1-K_{lk})
  +\frac{1}{x_{l}+x_{k}}(1-P_{l}P_{k}K_{lk})\right), \nonumber\\
  x_{l}, & & \nonumber \\
  D_{l}^{(B)} & = & x_{l}\nabla_{l}^{(B)},
\end{eqnarray}
where $K_{lk}$ and $P_{l}$ are elements of the $B_{N}$-type Weyl group, 
$K_{lk}$ is the coordinate exchange operator whose action is same as in the 
$A_{N-1}$ case and $P_{l}$ is the reflection operator whose action on 
multivariable functions is defined by
\begin{equation}
  (P_{l}f)(\cdots,x_{l},\cdots)=f(\cdots,-x_{l},\cdots).
\end{equation}
Commutation relations among the operators are given by 
\begin{eqnarray}
  \label{B-S-commutation}
  \bigl[\nabla_{l}^{(B)},\nabla_{m}^{(B)}\bigr]=0, 
  \quad \quad \bigl[x_{l},x_{m}\bigr]=0, \nonumber \\
  \bigl[\nabla_{l}^{(B)},x_{m}\bigr]
  =\delta_{lm}(1+a\sum_{\stackrel{\scriptstyle k=1}{k\neq m}}^{N}
  (1+P_{m}P_{k})K_{mk}+2b P_{m})
  \nonumber \\
  \quad \quad \quad \quad \quad \quad 
  -a(1-\delta_{lm})(1-P_{l}P_{m})K_{lm}, \nonumber \\
  \bigl[D_{l}^{(B)},D_{m}^{(B)}\bigr]
   =a(D_{m}^{(B)}-D_{l}^{(B)})(1+P_{l}P_{m})K_{lk},\nonumber \\
  \nabla_{l}^{(B)}\cdot 1=0,
\end{eqnarray}
and the commuting conserved operators are 
\begin{equation}
  \label{B-S-conserved}
   L_{k}^{(B)}=\sum_{l=1}^{N}(D_{l}^{(B)})^{k}
   \Bigr|_{\rm Sym,Even}, \quad [L_{k}^{(B)},L_{m}^{(B)}]=0,
   \quad k,m=1,2,\cdots,N,
\end{equation}
where the symbol $\Bigr|_{\rm Sym,Even}$
denotes the restriction of the operand to symmetric functions with
even parity.
We note that the restriction of the operands of the Dunkl operators
$D_{l}^{(B)}$ to even functions $\Bigr|_{\rm Even}$ yields
\begin{equation}
  \label{rest-Dunkl}
  D_{l}^{(B)}\Bigr|_{\rm Even}=x_{l}\frac{\partial}{\partial x_{l}}
   +a\sum_{\stackrel{\scriptstyle k=1}{k\neq l}}^{N}
   \frac{2x_{l}^{2}}{x_{l}^{2}-x_{k}^{2}}(1-K_{lk}).
\end{equation}
Comparing (\ref{rest-Dunkl}) with (\ref{A-S-Dunkl}), 
we notice that $D_{l}^{(B)}\Bigr|_{\rm Even}$ is equivalent to 
$2D_{l}^{(A)}$ with the change of variables,
$x_{l}\rightarrow x_{l}^{2}/2$. As a consequence, the symmetric
simultaneous eigenfunctions of the conserved operators $L_{k}^{(B)}$
with even parity are given by
the Jack symmetric polynomials whose 
arguments $x_{l}$ are replaced with $x_{l}^{2}/2$,
$J_{\lambda}(x_{1}^{2}/2,\cdots,x_{N}^{2}/2)
=2^{-|\lambda|}J_{\lambda}(\vecvar{x}^{2})$.
They form the orthogonal basis of the $B_{N}$-Sutherland model.

The Dunkl operators for the $B_{N}$-Calogero model are
\begin{eqnarray}
  \label{B-C-Dunkl}
  \alpha_{l}^{(B)}=\nabla_{l}^{(B)},\nonumber\\
  \alpha_{l}^{(B)\dagger}=-\frac{1}{2\omega}\nabla_{l}^{(B)}+x_{l},
  \nonumber\\
  d_{l}^{(B)}=\alpha_{l}^{(B)\dagger}\alpha_{l}^{(B)}.
\end{eqnarray}
Commutation relations among these operators are given by
\begin{eqnarray}
  \label{B-C-commutation}
  \bigl[\alpha_{l}^{(B)},\alpha_{m}^{(B)}\bigr]=0, 
  \quad \quad 
  \bigl[\alpha_{l}^{(B)\dagger},\alpha_{m}^{(B)\dagger}\bigr]=0, 
  \nonumber\\
  \bigl[\alpha_{l}^{(B)},\alpha_{m}^{(B)\dagger}\bigr]
  =\delta_{lm}(1+a\sum_{\stackrel{\scriptstyle k=1}{k\neq m}}^{N}
  (1+P_{m}P_{k})K_{mk}+2b P_{m})
  \nonumber \\
  \quad \quad \quad \quad \quad \quad \quad 
  -a(1-\delta_{lm})(1-P_{l}P_{m})K_{lm}, \nonumber\\
  \bigl[d_{l}^{(B)},d_{m}^{(B)}\bigr]
  =a(d_{m}^{(B)}-d_{l}^{(B)})(1+P_{l}P_{m})K_{lm},\nonumber \\ 
  \alpha_{l}^{(B)}\cdot 1=0,
\end{eqnarray}
and the mutually commuting conserved operators are 
\begin{equation}
  \label{B-C-conserved}
  I_{k}^{(B)}=\sum_{l=1}^{N}(d_{l}^{(B)})^{k}
  \Bigr|_{\rm Sym,Even},\quad [I_{k}^{(B)},I_{m}^{(B)}]=0,\quad
  k,m=1,2,\cdots,N.
\end{equation}
In a similar way to the translation between the $A_{N-1}$-Calogero and
Sutherland models, the simultaneous eigenfunctions of the above
conserved operators are obtained by putting 
$(\alpha_{i}^{(B)\dagger})^{2}/2$ into the arguments of the Jack
polynomials, $2^{-|\lambda|}l_{\lambda}(\vecvar{x})
=J_{\lambda}((\alpha_{1}^{\dagger})^{2}/2,\cdots,
(\alpha_{N}^{\dagger})^{2}/2)\cdot 1
=2^{-|\lambda|}J_{\lambda}(\vecvar{(\alpha}^{\dagger})^{2})\cdot 1$,
which form the orthogonal basis of the $B_{N}$-Calogero model.


\section{Cancellation of essential singularities}
\label{sec:singularity}
We show how the transformation 
(\ref{transform4}) of the non-symmetric number states causes
the essential singularity and how we can escape from it.
We note that the transformation of the number state is rewritten as
\begin{equation}
  \label{general-form}
  T|n_{1},\cdots,n_{N}\rangle
  =\e^{-\frac{1}{4\omega}{\cal O}_{{\rm L}}} 
  x_{1}^{n_{1}}x_{2}^{n_{2}} \cdots x_{N}^{n_{N}}. 
\end{equation}
First, we consider the $A_{N-1}$-Calogero model.
Action of Lassalle operator ${\cal O}_{{\rm L}}^{(A)}$ on a monomial
$x_{1}^{n_{1}} x_{2}^{n_{2}} \cdots x_{N}^{n_{N}}$ yields
\begin{eqnarray}
  \label{A-singular}
  \fl {\cal O}_{{\rm L}}^{(A)}
  x_{1}^{n_{1}}x_{2}^{n_{2}}\cdots x_{N}^{n_{N}}
  & = & \sum_{l=1}^{N}n_{l}(n_{l}-1)x_{1}^{n_{1}}\cdots
  x_{l}^{n_{l}-2}\cdots x_{N}^{n_{N}}\nonumber\\
  & & +a\sum_{\stackrel{\scriptstyle l,m=1}{l\neq m}}^{N}
  \frac{1}{x_{l}-x_{m}}(n_{l} x_{m}-n_{m} x_{l})
  x_{1}^{n_{1}}\cdots x_{l}^{n_{l}-1}
  \cdots x_{m}^{n_{m}-1}\cdots x_{N}^{n_{N}}.
\end{eqnarray}
As we see, action of the Lassalle operator generally generates
poles at $x_{l}=x_{m}$ in the second term.
The action of the exponentiation of the
Lassalle operator means the action of the Lassalle operator infinitely
many times, which develops such poles into essential singularities.
We can remove such poles by symmetrization.
Acting the Lassalle operator ${\cal O}_{{\rm L}}^{(A)}$
on a symmetrized monomial
$m_{\lambda}(\mbox{\boldmath $x$})$ (\ref{mono-sym-func}), we have
\begin{eqnarray}
  \label{A-non-singular}
  \fl {\cal O}_{{\rm L}}^{(A)} m_{\lambda}
  =a\!\sum_{\stackrel{\scriptstyle l,m=1}{l\neq m}}^{N}
  \sum_{\stackrel{\scriptstyle \sigma\in S_{N}}{\rm distinct}}
  \frac{\lambda_{\sigma(l)} 
    x_{l}^{\lambda_{\sigma(l)}-1} 
    x_{m}^{\lambda_{\sigma(m)}}
    -\lambda_{\sigma(m)}
    x_{l}^{\lambda_{\sigma(l)}}
    x_{m}^{\lambda_{\sigma(m)}-1}}{x_{l}-x_{m}}
    x_{1}^{\lambda_{\sigma(1)}}\cdots^{\stackrel{\scriptstyle l}{\vee}}
    \cdots^{\stackrel{\scriptstyle m}{\vee}}\cdots
    x_{N}^{\lambda_{\sigma(N)}}\nonumber \\
   +(\mbox{some symmetrized monomials}) \nonumber \\
  \fl \hspace{14mm}
  =a\!\sum_{\stackrel{\scriptstyle l,m=1}{l\neq m}}^{N}
  \sum_{\stackrel{\scriptstyle \sigma\in S_{N}}{\rm distinct}}
  \lambda_{\sigma(l)}
   \frac{x_{l}^{\lambda_{\sigma(l)}-1}x_{m}^{\lambda_{\sigma(m)}}
        -x_{m}^{\lambda_{\sigma(l)}-1}x_{l}^{\lambda_{\sigma(m)}}}
        {x_{l}-x_{m}}
        x_{1}^{\lambda_{\sigma(1)}}\cdots^{\stackrel{\scriptstyle l}{\vee}}
     \cdots^{\stackrel{\scriptstyle m}{\vee}}
     \cdots x_{N}^{\lambda_{\sigma(N)}} \nonumber\\
     +(\mbox{some symmetrized monomials})\nonumber\\
  \fl \hspace{14mm} = (\mbox{some symmetrized monomials}), 
\end{eqnarray}
where $ ^{\stackrel{\scriptstyle l}{\vee}}$ denotes a missing $x_{l}$.
We have used the fact
that the numerator in the second expression has a factor $(x_{l}-x_{m})$,
which cancels the denominator out.
Thus we have removed poles by symmetrization.

In a similar way, essential singularities appear in the $B_{N}$ case.
We act the $B_{N}$-Lassalle operator ${\cal O}_{{\rm L}}^{(B)}$
on a monomial $x_{1}^{n_{1}} x_{2}^{n_{2}} \cdots x_{N}^{n_{N}}$ and get
\begin{eqnarray}
  \label{B-singular1}
  \fl {\cal O}_{{\rm L}}^{(B)}x_{1}^{n_{1}} x_{2}^{n_{2}} \cdots x_{N}^{n_{N}}
  = \sum_{l=1}^{N}\bigl(n_{l}(n_{l}-1)+2bn_{l}\bigr)x_{1}^{n_{1}}\cdots
  x_{l}^{n_{l}-2}\cdots x_{N}^{n_{N}}\nonumber\\
  \fl \Usr\Usr\Usr\Usr\Usr
  +2a\sum_{\stackrel{\scriptstyle l,m=1}{l\neq m}}^{N}
   \frac{n_{l}-n_{m}}{x_{l}^{2}-x_{m}^{2}}
   x_{1}^{n_{1}} x_{2}^{n_{2}} \cdots x_{N}^{n_{N}}.
\end{eqnarray}
The second term yields poles at $x_{l}=x_{m}$ and
$x_{l}=-x_{m}$. In addition to those, the first term also yields
a pole at $x_{l}=0$ when $n_{l}=1$. Considering successive actions
of the Lassalle operator, we conclude that the poles of the second type
appear when the powers $n_{l}$ are odd. Thus we have to restrict the 
operand to even functions. Symmetrizing \myref{B-singular1}, we
act the Lassalle operator on an even symmetrized monomial,
\begin{eqnarray}
  \label{B-singular2}
  \fl {\cal O}_{{\rm L}}^{(B)}m_{\lambda}
      =2a\sum_{\stackrel{\scriptstyle l,m=1}{l\neq m}}^{N}
       \sum_{\stackrel{\scriptstyle \sigma\in S_{N}}{\rm distinct}}
      \frac{\lambda_{\sigma(l)}-\lambda_{\sigma(m)}}{x_{l}^{2}-x_{m}^{2}}
       x_{1}^{\lambda_{\sigma(1)}}\cdots
       x_{N}^{\lambda_{\sigma(N)}}\nonumber\\
  \fl \hspace{16mm} +(\mbox{some even symmetrized monomials})\nonumber \\
  \fl \hspace{14mm}
      =2 a \sum_{\stackrel{\scriptstyle l,m=1}{l\neq m}}^{N}
      \sum_{\stackrel{\scriptstyle \sigma\in S_{N}}{\rm distinct}}
      \lambda_{\sigma(l)}
      \frac{x_{l}^{\lambda_{\sigma(l)}}x_{m}^{\lambda_{\sigma(m)}}
            -x_{m}^{\lambda_{\sigma(l)}}x_{l}^{\lambda_{\sigma(m)}}}
            {x_{l}^{2}-x_{m}^{2}}
       x_{1}^{\lambda_{\sigma(1)}}\cdots^{\stackrel{\scriptstyle l}{\vee}} 
       \cdots^{\stackrel{\scriptstyle m}{\vee}} 
       \cdots x_{N}^{\lambda_{\sigma(N)}}\nonumber\\
 \fl \hspace{16mm} +(\mbox{some even symmetrized monomials})\nonumber\\
 \fl \hspace{14mm} = (\mbox{some even symmetrized monomials}).
\end{eqnarray}
We have used the fact that the numerator in the second expression has a
factor $(x_{l}^{2}-x_{m}^{2})$ when $\lambda_{\sigma(l)}$ and
$\lambda_{\sigma(m)}$ are even.
Consequently, we can escape from the essential singularity by restricting
the operands of the $B_{N}$-Lassalle operator to symmetric function with
even parity.


\section{Explicit forms of new orthogonal bases}
\label{sec:explicit_forms}
We show explicit forms of some of the new orthogonal
bases and variants of the Jack polynomial. They are expressed
in terms of the monomial symmetric functions.

The first seven of the new orthogonal symmetric polynomials 
for the $A_{N-1}$-Calogero model are  
\begin{eqnarray}
 M_{0}=m_{0}=1,\quad
 M_{1}=m_{1},\nonumber \\
 M_{2}=m_{2}-\frac{1}{2\omega}N\bigl[(N-1)a+1\bigr]m_{0},\nonumber \\
 M_{1^{2}}=m_{1^{2}}+\frac{1}{4\omega}N(N-1)a\,m_{0},\nonumber \\
 M_{3}=m_{3}-\frac{3}{2\omega}\bigl[(N-1)a+1\bigr]m_{1},\nonumber \\
 M_{2,1}=m_{2,1}-\frac{1}{2\omega}(N-1)\bigl[(N-3)a+1\bigr]m_{1},
 \nonumber \\
 M_{1^{3}}=m_{1^{3}}+\frac{1}{4\omega}(N-1)(N-2)a\,m_{1},
 \label{example-A}
\end{eqnarray}
and the first four of those for the $B_{N}$-Calogero model are
\begin{eqnarray}
 Y_{0}=m_{0}=1,\nonumber \\
 Y_{1}=m_{1}-\frac{1}{2\omega}N\bigl[2(N-1)a+2b+1\bigr]m_{0},\nonumber \\
 Y_{2}=m_{2}-\frac{1}{\omega}\bigl[4(N-1)a+2b+3\bigr]m_{1} \nonumber\\
 \quad \quad  +\frac{1}{4\omega^{2}}N\bigl[4(N-1)a+2b+3\bigr]
                              \bigl[2(N-1)a+2b+1\bigr]m_{0},\nonumber \\
 Y_{1^{2}}=m_{1^{2}}-\frac{1}{2\omega}(N-1)\bigl[2(N-2)a+2b+1\bigr]m_{1}
 \nonumber\\
 \quad \quad
 +\frac{1}{8\omega^{2}}N(N\!-\!1)\bigl[2(N\!-2)a+2b+1\bigr]
                             \bigl[2(N\!-\!1)a+2b+1\bigr]m_{0}.
  \label{example-B}
\end{eqnarray}

We also present
the first seven of the Hi-Jack polynomials,
\begin{eqnarray}
  \label{Hi-Jack-poly}
  j_{0}=M_{0}=1,\quad
  j_{1}=M_{1}=m_{1},\nonumber \\
  j_{2}=M_{2}+\frac{2a}{a+1}M_{1^{2}}\nonumber \\
  \hspace{6mm}
              =m_{2}+\frac{2a}{a+1}m_{1^{2}}
               -\frac{1}{2\omega}\frac{N(Na+1)}{a+1}m_{0},\nonumber \\
  j_{1^{2}}=M_{1^{2}},\nonumber \\
  j_{3}=M_{3}+\frac{3a}{a+2}M_{2,1}
               +\frac{6a^{2}}{(a+1)(a+2)}M_{1^{3}}\nonumber \\
  \hspace{6mm}
   =m_{3}+\frac{3a}{a+2}m_{2,1}+\frac{6a^{2}}{(a+1)(a+2)}m_{1^{3}}
  \nonumber \\
  \hspace{9mm}
    -\frac{3}{2\omega}\frac{(Na+1)(Na+2)}{(a+1)(a+2)}m_{1},\nonumber \\
  j_{2,1}=M_{2,1}+\frac{6a}{2a+1}M_{1^{3}}\nonumber \\
  \hspace{8mm}
   =m_{2,1}+\frac{6a}{2a+1}m_{1^{3}}
    +\frac{1}{2\omega}\frac{(Na-1)(Na+1)(a-1)}{2a+1}m_{1},\nonumber \\
  j_{1^{3}}=M_{1^{3}},
\end{eqnarray}
and the first four of the multivariable Laguerre polynomials,
\begin{eqnarray}
  l_{0}=Y_{0}=1, \quad
  l_{1}=Y_{1}, \nonumber\\
  l_{2}=Y_{2}+\frac{2a}{a+1}Y_{1^{2}} \nonumber \\
  \hspace{5mm}
   =m_{2}+\frac{2a}{a+1}m_{1^{2}}\nonumber\\
  \hspace{9mm}
    -\frac{1}{\omega}\frac{1}{a+1}
    \bigl[(N-1)(2Na+2b+5)a+(a+1)(2b+3)\bigr]m_{1}\nonumber\\
  \hspace{9mm}
    +\frac{1}{4\omega^{2}}\frac{1}{a+1}
    N\bigl[2(N-1)a+2b+1\bigr]\nonumber \\ 
  \hspace{9mm}\times
     \bigl[(N-1)(2Na+2b+5)a+(a+1)(2b+3)\bigr]m_{0}, \nonumber\\
  l_{1^{2}}=Y_{1^{2}}.
\end{eqnarray}


\end{document}